\documentclass[twocolumn,floats,floatfix,superscriptaddress,aps,pra]{revtex4}
\usepackage{amsfonts}
\usepackage{amssymb}
\usepackage{amsmath}
\usepackage{calc}
\usepackage{graphicx}
\usepackage{bm}
\usepackage{makecell}
\usepackage{booktabs}

\def\be{ \begin{equation} }
\def\ee{ \end{equation} }
\def\bea{ \begin{eqnarray} }
\def\eea{ \end{eqnarray} }
\def\bse{ \begin{subequations} }
\def\ese{ \end{subequations} }
\def\ba{ \begin{array} }
\def\ea{ \end{array} }

\def\to{\rightarrow}

\def\U{\mathbf{U}}

\def\be{ \begin{equation} }
\def\ee{ \end{equation} }
\def\bea{ \begin{eqnarray} }
\def\eea{ \end{eqnarray} }
\def\bse{ \begin{subequations} }
\def\ese{ \end{subequations} }

\def\U{\mathbf{U}}

\def\to{\rightarrow}

\newcommand{\ket}[1]{\vert #1\rangle}

\begin{document}

\author{Boyan T. Torosov}
\affiliation{Institute of Solid State Physics, Bulgarian Academy of Sciences, 72 Tsarigradsko chauss\'{e}e, 1784 Sofia, Bulgaria}
\author{Michael Drewsen}
\affiliation{Department of Physics and Astronomy, Aarhus University, DK-8000 Aarhus C, Denmark}
\author{Nikolay V. Vitanov}
\affiliation{Department of Physics, St Kliment Ohridski University of Sofia, 5 James Bourchier blvd, 1164 Sofia, Bulgaria}

\title{Chiral resolution by composite Raman pulses}

\date{\today}

\begin{abstract}
We present two methods for efficient detection of chiral molecules based on sequences of single pulses and Raman pulse pairs. The chiral molecules are modelled by a closed-loop three-state system with different signs in one of the couplings for the two enantiomers. One method uses a sequence of three interaction steps: a single pulse, a Raman pulse, and another single pulse. The other method uses a sequence of only two interaction steps: a Raman pulse, and a single pulse. The second method is simpler and faster but requires a more sophisticated Raman pulse than the first one.
Both techniques allow for straightforward generalizations by replacing the single and Raman pulses with composite pulse sequences. The latter achieve very high signal contrast and far greater robustness to experimental errors than by using single pulses. We demonstrate that both constant-rotation (i.e., with phase compensation) and variable-rotation (i.e., with phase distortion) composite pulses can be used, the former being more accurate and the latter being simpler and faster. 
\end{abstract}

\maketitle

\section{Introduction\label{Sec:intro}}

In physics, symmetry (and asymmetry) is vital to understanding and predicting the phenomena in the surrounding world. Symmetry has always played an important role in the description of how our universe works, but it has become quintessential after the formulation and proof of Emmy Noether's theorem \cite{Noether}, which provides the link between symmetry and conservation laws.
As a special type of asymmetry, chirality is of crucial significance in many branches of contemporary science, e.g. in chemistry, biotechnologies, and pharmaceutics.
A chiral molecule, also called an enantiomer, is one that cannot be superimposed on its mirror image by translation and rotation.
Such molecule pairs have identical physical properties, neglecting the small differences due to the electroweak interaction \cite{electroweak}. Nevertheless, the chemical properties of the two enantiomers may be entirely different.
This is essential in pharmaceutics, where the chiral purity of a particular substance may be crucial to the drug efficiency.

Traditionally, enantiomer detection and separation is based on slow, complicated, and expensive chemical techniques, such as crystallization, derivatization, kinetic resolution, and chiral chromatography \cite{Ahuja}.
Alternatively, one may use chiroptical spectroscopy to break the symmetry of the enantiomers by interaction with circularly polarized light \cite{Chiroptical}.
Some prevalent chiroptical methods are optical rotary dispersion \cite{Chiroptical}, circular dichroism \cite{dichroism}, vibrational circular dichroism \cite{Nafie,NafieStephens}, and Raman optical activity \cite{Nafie,Barron}.
These methods rely on the magnetic-dipole interaction between the circularly polarized light and the molecules.
Furthermore, methods based on linearly polarized light have been developed, using the much stronger electric-dipole interaction \cite{LinearLight}.
These methods make use of the sign difference of some of the transition dipole moments of the two enantiomers, which is then mapped onto population differences by using quantum systems with three or four states driven in closed-loop interaction schemes, thereby creating interferometric linkages.
Another approach has been developed by Shapiro and co-workers \cite{Shapiro}, who used concepts from adiabatic passage methods \cite{STIRAP} to detect and separate enantiomers.
Finally, rotational spectroscopy has been used to develop methods, such as microwave three-wave mixing (M3WM) \cite{M3WM, Li2008, Hirota} for chiral analysis in gas-phase samples.

Over the last few years there is a growing interest on the topic. Some remarkable results are the determination of enantiomeric excess, based on chirality-dependent AC Stark effects \cite{Li2019, LehmannACstark}, a theory on the enantiomeric separation in the presence of spatial degeneracy \cite{Lobsiger2015, Lehmann2018} and a detailed study of the enantioselective three-wave mixing spectroscopy, where it was shown,	by using group theoretical arguments, that three mutually orthogonal polarizations are needed to achieve chiral resolution \cite{Koch,Goetz2019}.
Also, a simple and fast method for chiral resolution, based on shortcuts to adiabaticity \cite{STA}, has been proposed \cite{VitanovDrewsen}.

Recently, we developed a method for optical detection of chiral molecules, based on simple sequences of resonant pulses \cite{TorosovDrewsenVitanov}. These allowed for a robust and high-fidelity optimization using composite pulses (CPs).
A CP is a sequence of pulses with appropriately chosen relative phases.
These phases are used as control parameters to shape the excitation profile in a desired fashion.
In such a way one can produce broadband, narrowband and passband profiles, robust coherent superpositions, optimized adiabatic techniques, high-fidelity quantum gates, etc.

In the current work, we develop this approach further, by using Raman pulses in addition to the pulses directly coupling a single transition.
Raman pulses have been used before for chiral resolution, e.g. in \cite{Li2008}, where they are applied on resonance, or in \cite{Wu2020}, where they are applied off-resonance in order to adiabatically eliminate the upper state.
Our focus now is to use resonant Raman pulses, which allow to benefit from the powerful technique of composite pulses, in order to achieve efficient and robust chirality-dependent population transfer.

To this end, we model the chiral molecules as a delta-type system, where one of the three couplings, marked as $P,S,Q$, differs in sign in the two enantiomers, see Fig.~\ref{fig:scheme}.
In Sec.~\ref{Sec:method} we explain our method and show how it can map this sign difference onto population transfer to different states.
We also discuss in details the specific pulse sequences, which can be used for the population transfer.
In Section~\ref{Sec:CompositePulses} we  demonstrate how constant-rotation composite pulses can be used to improve the contrast and robustness of the method, and Sec.~\ref{Sec:CompositePulses-variable} demonstrates similar features with the simpler and faster variable-rotation composite pulses.
Finally, the conclusions are summarized in Sec.~\ref{Sec:conclusions}.


\section{Description of the method}\label{Sec:method}

\begin{figure}[tb]
	\includegraphics[width=8.5cm]{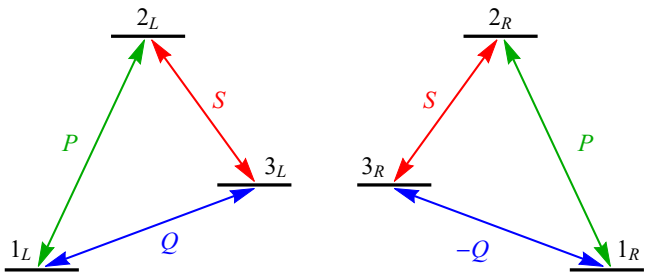}
	\caption{
		Coupling schemes for molecules with left and right handedness.
	}
	\label{fig:scheme}
\end{figure}

We now describe our method to produce chirality-dependent population transfer, based on two types of sequences, both using a combination of single and Raman pulses.
We model the chiral molecules as a delta-type system, which is illustrated by the energy level scheme in Fig.~\ref{fig:scheme}.
We assume that the two enantiomers differ only by the sign  of the $\ket{1}\leftrightarrow \ket{3}$ coupling.
The first type of sequences consists of a single pulse, followed by a Raman pair of pulses, followed by another single pulse.
Such sequence has been used before, e.g. in Ref.~\cite{Li2008}.
In this work we consider also other similar sequences, and, more importantly, propose an optimization by using composite pulses.
The second type uses a sequence of just one Raman pulse pair and one single pulse.
Here we study these sequences in details, and then we examine how one can optimize them by using composite pulses.
In all our derivations, we shall assume that the system is initially in state $\ket{1}$.

\subsection{Single-Raman-single sequence}

\begin{table}
	\begin{tabular}{c c c}
		\hline
		\addlinespace[1ex]
		\vspace{2pt}
		pulse sequence & final & final \\
		 & state ($L$) & state ($R$)  \\
		\hline
		\addlinespace[1ex]
			$Q(\tfrac{\pi}{2})\left[ P(\tfrac{\pi}{\sqrt{2}}),iS(\tfrac{\pi}{\sqrt{2}})\right]Q(\tfrac{\pi}{2})$ & 2 & 3 \\
		\addlinespace[1ex]
		$Q(\tfrac{\pi}{2})\left[ P(\tfrac{\pi}{\sqrt{2}}),iS(\tfrac{\pi}{\sqrt{2}})\right]-Q(\tfrac{\pi}{2})$ & 2 & 1 \\
		\addlinespace[1ex]
		$-Q(\tfrac{\pi}{2})\left[ P(\tfrac{\pi}{\sqrt{2}}),iS(\tfrac{\pi}{\sqrt{2}})\right]Q(\tfrac{\pi}{2})$ & 1 & 2 \\
		\addlinespace[1ex]
		$-Q(\tfrac{\pi}{2})\left[ P(\tfrac{\pi}{\sqrt{2}}),iS(\tfrac{\pi}{\sqrt{2}})\right]-Q(\tfrac{\pi}{2})$ & 3 & 2 \vspace{2pt} \\
		\hline
		\addlinespace[1ex]
		$\left[P(\xi_1\pi),iS(\xi_2\pi)\right]Q(\tfrac{\pi}{2})$ & 3 & 1 \\
		\addlinespace[1ex]
		$\left[P(\xi_2\pi),iS(\xi_1\pi)\right]Q(\tfrac{\pi}{2})$ & 1 & 3 \vspace{2pt} \\
		\hline
	\end{tabular}
	\caption{
		Sequences of pulses and corresponding transition probabilities for the $L$ and $R$ enantiomers. The numbers in the round brackets correspond to the pulse area of the pulse of the corresponding $P$, $S$ or $Q$ transitions, including a specific phase, if applicable, with $\xi_1=\sqrt{2+\sqrt{2}}$ and $\xi_2=\sqrt{2-\sqrt{2}}$.
The square brackets indicate simultaneous (Raman) pulses.
	}
	\label{Table:sequences}
\end{table}

\begin{figure}[tb]
	\includegraphics[width=8.5cm]{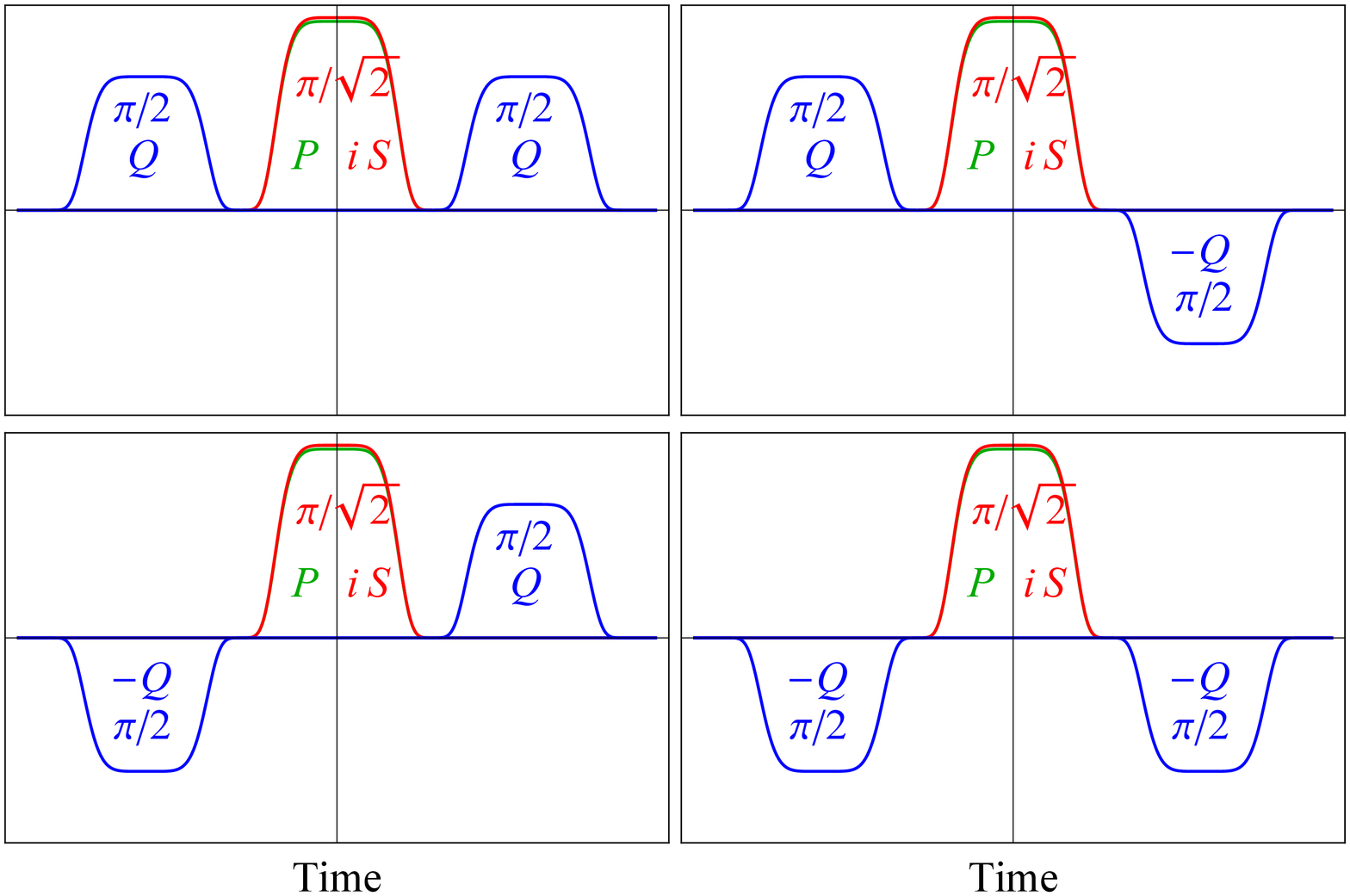}
	\caption{
		Examples of sequences of two $\pi/2$ pulses with a Raman pulse in-between, leading to chiral-dependent population transfer.
The numbers $\pi/2$ and $\pi/\sqrt{2}$ indicate the temporal area on the respective pulse.
	}
	\label{fig:SequenceThreeRaman}
\end{figure}

The \emph{single-Raman-single sequence} consists of three sequential separated interactions.
The first step is to apply a single $\pi/2$ pulse on the $Q$ transition.
This $\pi/2$ rotation corresponds to the following propagator,
\be\label{prop-iQ}
\U_{Q} = \left[ \begin{array}{ccc}   \frac{1}{\sqrt{2}} & 0 & \frac{\mp i}{\sqrt{2}}  \\ 0 & 1 & 0 \\  \frac{\mp i}{\sqrt{2}} & 0 & \frac{1}{\sqrt{2}}  \end{array}\right] ,
\ee
where the $\mp$ sign corresponds to the $L$ and $R$ chiral molecules.
 This transfers the system, initially in state $\ket{1}$, to states $\frac{1}{\sqrt{2}}\left( \ket{1} \mp i\ket{3} \right)$.
 In the second step we apply a Raman interaction, which consists of two simultaneous $P$ and $S$ pulses, each of area equal to $\pi/\sqrt{2}$, where the $S$ pulse has a phase of $\pi/2$ relative to the $P$ pulse.
 The propagator for this step is
\be
\U_{[P,iS]}=\left[ \begin{array}{ccc} \frac{1}{2} & \frac{-i}{\sqrt{2}} & \frac{-i}{2} \\ \frac{-i}{\sqrt{2}} & 0 & \frac{1}{\sqrt{2}}  \\ \frac{i}{2} & \frac{-1}{\sqrt{2}} & \frac{1}{2}  \end{array}\right],
\ee
which transfers the system into states $-i\ket{2}$ and $\frac{1}{\sqrt{2}}\left( \ket{1} + i\ket{3} \right)$ for the $L$ and $R$ chiralities, respectively. Finally, the third step is identical to the first one.
As seen from Eq.~\eqref{prop-iQ}, the $L$-handed system stays in state $-i\ket{2}$, while the $R$-handed one is transferred to state $i\ket{3}$.
The total propagator of this sequence is $\U = \U_{Q} \U_{[P,iS]} \U_{Q}$, which, after some trivial calculations, becomes
\be
\U^{(L)} = \left[ \begin{array}{ccc} 0 & 0 & -i \\-i & 0 & 0  \\ 0 & -1 & 0  \end{array}\right],\quad
\U^{(R)} = \left[ \begin{array}{ccc} 0 & -i & 0 \\0 & 0 & 1  \\ i & 0 & 0  \end{array}\right].
\ee
Therefore, chiral resolution is achieved by using this three-steps procedure.
We can write this sequence as
\be\label{RamanEqxample}
Q(\tfrac{\pi}{2})\left[ P(\tfrac{\pi}{\sqrt{2}}),iS(\tfrac{\pi}{\sqrt{2}})\right]Q(\tfrac{\pi}{2}),
\ee
where the pulses in the square brackets constitute the Raman interaction.
Other similar sequences with different signs of the $Q$ pulses, leading to chiral resolution, are
 listed in Table.~\ref{Table:sequences} (top) and illustrated in Fig.~\ref{fig:SequenceThreeRaman}. In the next subsection we describe an even simpler procedure, consisting of just two steps, for achieving our goal of chiral-dependent population transfer.

We note that we have chosen the $\pi/2$ phase shift to be attached to the $S$ field. However, chiral resolution can be achieved if it is attached to one of the other two fields, $P$ or $Q$.

\subsection{Raman-Single sequence}

\begin{figure}[tb]
	\includegraphics[width=8.5cm]{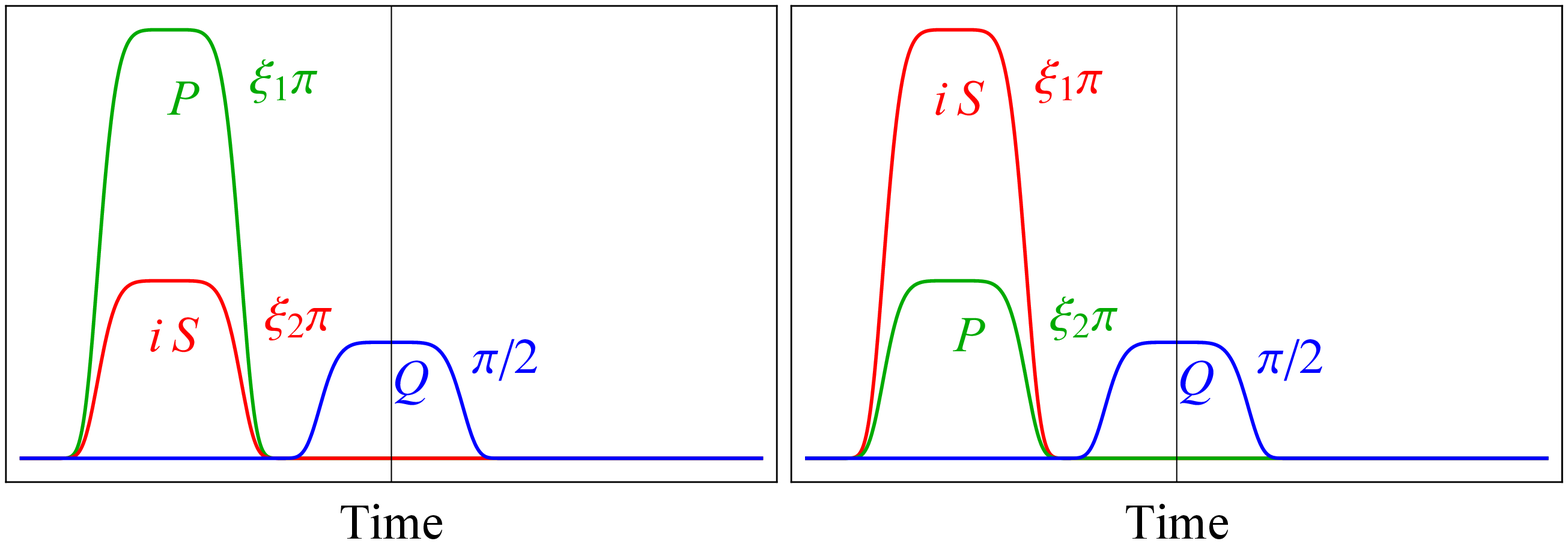}
	\caption{
		Examples of sequences of a Raman pulse and a $\pi/2$ resonant pulse after it, leading to chiral-dependent population transfer.
The $Q$ pulse has a temporal area of $\pi/2$.
On the left, the $P$ pulse has an area of $\xi_1\pi$, while the $S$ pulse has an area of $\xi_2\pi$, and a relative phase of $\pi/2$, with $\xi_1=\sqrt{2+\sqrt{2}}$ and $\xi_2=\sqrt{2-\sqrt{2}}$.
On the right, the $P$ pulse has an area of $\xi_2\pi$, while the $S$ pulse has an area of $\xi_1\pi$, still with a relative phase of $\pi/2$.
	}
	\label{fig:SequenceTwoRaman}
\end{figure}

The \emph{Raman-single sequence} is made of the following two steps.
First, we apply a Raman pulse, consisting of two simultaneous resonant pulses on the $P$ and $S$ transitions, where the $P$ pulse has an area of $\xi_1\pi$, while the $S$ pulse has an area of $\xi_2\pi$, and a relative phase of $\pi/2$. Here $\xi_1=\sqrt{2+\sqrt{2}}$ and $\xi_2=\sqrt{2-\sqrt{2}}$. The propagator for this step is
\be
\U_{[P,iS]} = \left[ \begin{array}{ccc}   \frac{-1}{\sqrt{2}} & 0 & \frac{-i}{\sqrt{2}}  \\ 0 & -1 & 0 \\  \frac{i}{\sqrt{2}} & 0 & \frac{1}{\sqrt{2}}  \end{array}\right],
\ee
and therefore, the system, which initially was in state $\ket{1}$, is now transferred to state $\frac{-1}{\sqrt{2}}(\ket{1} - i\ket{3})$. During the second step we apply a single $\pi/2$-pulse on the $Q$ transition.
The propagator for this step is given by Eq.~\eqref{prop-iQ}, which, as can be easily calculated, transfers the system into $i\ket{3}$ for the $L$ chirality and $-\ket{1}$ for the $R$ chirality. Explicitly, the final propagators are
\be
\U^{(L)} = \left[ \begin{array}{ccc} 0 & 0 & -i \\0 & -1 & 0  \\ i & 0 & 0  \end{array}\right],\quad
\U^{(R)} = \left[ \begin{array}{ccc} -1 & 0 & 0 \\0 & -1 & 0  \\ 0 & 0 & 1  \end{array}\right].
\ee
Hence, we can achieve chiral resolution of the two enantiomers by a sequence of only one Raman pulse and one single pulse.
A similar sequence, where the pulse areas of the $P$ and $S$ pulses are interchanged, can also be used.
These are schematically illustrated in Fig.~\ref{fig:SequenceTwoRaman}, and the resulting population transfer is shown in Table~\ref{Table:sequences} (bottom).

In the next section, we describe how we can use composite pulses to improve the described procedures and achieve very high efficiency and robustness to experimental errors.

\section{Optimization by composite pulses}\label{Sec:CompositePulses}

\subsection{Composite single-Raman-single sequences}

We now show how our approach can be made robust to errors by replacing the single and Raman pulses with composite sequences.
First, we study the {single-Raman-single} method.
To be specific, we look at the sequence \eqref{RamanEqxample}, consisting of two single $Q$ pulses and a Raman
$[P,iS]$ pulse in-between.

There are generally two types of CPs.
The first type are the so-called \emph{variable rotations}, for which the moduli of the propagator elements (i.e. the square roots of probabilities) are robust to errors in the experimental parameters  but the phases of the propagator are not.
In the second type of CPs, which are called \emph{constant rotations}, both the moduli \emph{and} the phases of the propagator elements are robust to errors.
Clearly, we need to use the second type for our method in order to obtain robust excitation profiles with high fidelity because the precise phase relations are essential for its operation.

The shortest CP which offers constant $\pi/2$ rotation that can replace the $Q(\tfrac{\pi}{2})$ pulse is
\be\label{NVV1}
A'_{\phi_1} B_{\phi_2} A'_{\phi_1},
\ee
where $A'=0.6399\pi(1+\epsilon)$, $B=\pi(1+\epsilon)$ is a nominal $\pi$-pulse, and the pulse phases are $\phi_1=1.6558\pi$, $\phi_2=0.4413\pi$. This CP compensates pulse area errors up to order $O(\epsilon)$ and has a total nominal pulse area of about $2.28\pi$.
Here we have introduced the dimensionless parameter $\epsilon$, which is used in this work as a measure of the deviation from the perfect value of the corresponding pulse area.

Another symmetric constant-rotation $\pi/2$ CP is
\be\label{NVV2}
A''_{\phi_1} B_{\phi_2} B_{\phi_3} B_{\phi_2} A''_{\phi_1},
\ee
where $A''=0.45\pi(1+\epsilon)$, $\phi_1 = 1.4494\pi$, $\phi_2 = 0.0106\pi$, $\phi_3 = 0.8179\pi$.
It has a larger total nominal area of $3.9\pi$ but provides second-order error compensation $O(\epsilon^2)$.
The third example is the famous asymmetric BB1 sequence of Wimperis \cite{Wimperis},
\be\label{BB1}
A_0 B_{\chi} B_{3\chi} B_{3\chi} B_{\chi},
\ee
where $A=\pi(1+\epsilon)/2$ is a nominal $\pi/2$-pulse, and $\chi=\arccos(-1/8)\approx 0.5399\pi$.
It has a total area of $4.5\pi$ and offers second-order error compensation $O(\epsilon^2)$.

\begin{figure}[tb]
	\includegraphics[width=\columnwidth]{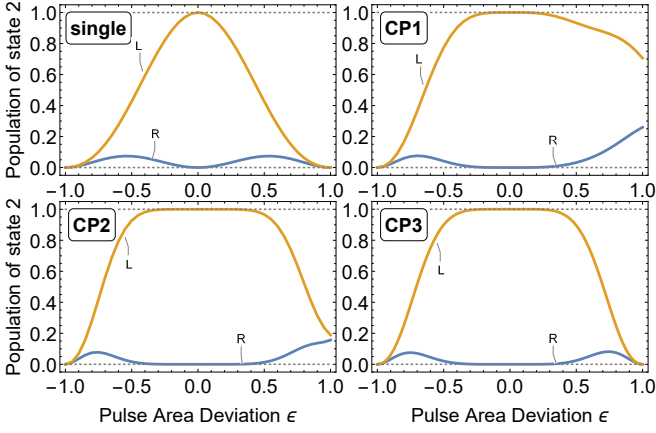}
	\caption{
		Population $P_2$ vs pulse area deviation for the single-Raman-single sequence (top left frame) and when the constant $\pi/2$ rotations from Eq.~\eqref{RamanHalfPi} are used to replace the single and Raman pulses. In frames CP1, CP2, and CP3 we have used the CPs \eqref{NVV1}, \eqref{NVV2}, and \eqref{BB1}, respectively.
	}
	\label{fig:vsAreaRaman3Wimperis}
\end{figure}

The $P$ and $S$ pulses in the Raman interaction should also be replaced by a composite sequence, which is a more demanding task because CPs have been developed primarily for two-state systems.
Nevertheless, CPs in multistate systems have been studied in the literature, and here we shall make use of the results obtained in \cite{CpMultiState, RamanGates}.
Namely, by using the Majorana decomposition \cite{MajoranaDecomposition} we map the initial three-state configuration to a much simpler two-state system.
By applying this approach, it is straightforward to substitute the single pulses with CPs and use the standard theory of two-state CPs \cite{CpMultiState,RamanGates}.
We can again use one of the constant $\pi/2$ rotations \eqref{NVV1}, \eqref{NVV2}, or \eqref{BB1}, to produce a composite Raman $\pi/\sqrt{2}$ pulse, where the $P$ and $S$ pulses in the Raman interaction $\left[P(\tfrac{\pi}{\sqrt{2}}),iS(\tfrac{\pi}{\sqrt{2}})\right]$ should be replaced with one of the following sequences,
\bse\label{RamanHalfPi}
\begin{align}
&\xi\pi/2 \to \left( \xi A' \right)_{\phi_1} \left(\xi \pi \right)_{\phi_2}\left( \xi A' \right)_{\phi_1}  ,\label{RamanHalfPiA}\\
&\xi\pi/2 \to \left( \xi A'' \right)_{\phi_1} \left(\xi \pi \right)_{\phi_2}\left(\xi \pi \right)_{\phi_3}\left(\xi \pi \right)_{\phi_2}\left( \xi A'' \right)_{\phi_1}  ,\label{RamanHalfPiB}\\
&\xi\pi/2 \to \left( \xi\pi/2 \right)_0 \left(\xi \pi \right)_{\chi} \left( \xi\pi \right)_{3\chi} \left( \xi\pi \right)_{3\chi} \left( \xi \pi\right)_{\chi}  ,\label{RamanHalfPiC}
\end{align}
\ese
with $\xi=\sqrt{2}$, corresponding to the CP sequences \eqref{NVV1}, \eqref{NVV2}, and \eqref{BB1}.

The performance of the method is illustrated in Fig.~\ref{fig:vsAreaRaman3Wimperis}, where we plot the population of state $\ket{2}$ as a function of the pulse area error.
The implementation with single pulses (top left) achieves chiral resolution in a narrow range around zero error ($\epsilon=0$).
Using composite pulses broaden this range considerably, even the shortest CP (top right) of Eq.~\eqref{NVV1}.
The longer five-pulse sequences of Eqs.~\eqref{NVV2} and \eqref{BB1} further broaden the high-contrast range.
In such a way, we obtain high-contrast chiral separation of the $L$ and $R$ molecules, which is very robust to experimental imperfections in the pulse area, which may derive from intensity fluctuations or improper pulse duration.

\subsection{Composite Raman-single sequences}

\begin{figure}[tb]
	\includegraphics[width=\columnwidth]{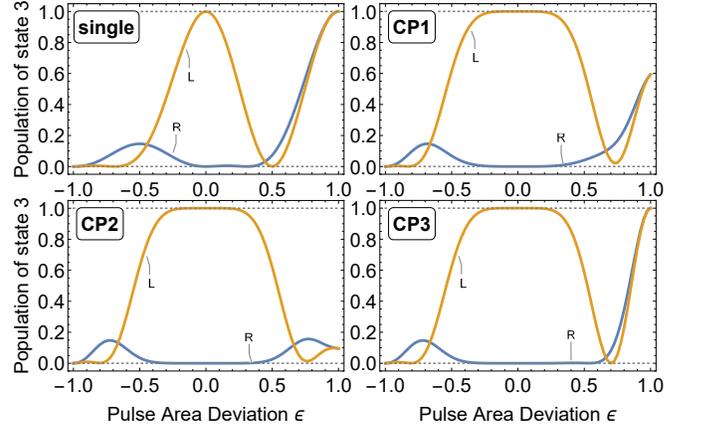}
	\caption{
		Population $P_3$ vs pulse area deviation for the Raman-single sequence (top left frame) and when composite sequences are used to replace the single pulse and the Raman pulse. The Raman pulse is replaced with the CP from Eq.~\eqref{2PiCP}, as shown in Eq.~\eqref{RamanTwoPi}, while the single $Q(\pi/2)$ pulse is replaced with the CPs \eqref{NVV1}, \eqref{NVV2}, and \eqref{BB1} (frames CP1, CP2, and CP3, respectively).
	}
	\label{fig:vsAreaRaman2Wimperis}
\end{figure}

Now we consider the Raman-single method and, to be specific, we study the sequence
\be\label{RamanSingleSequence}
\left[P(\xi_1\pi),iS(\xi_2\pi)\right] Q(\tfrac{\pi}{2}).
\ee
In order to replace the Raman pulses with composite sequences, in the case when the $P$ and $S$ pulses have different pulse areas, one can use the Morris-Shore (MS) transformation \cite{MS}. 
It has been shown in ref. \cite{RamanGates} that a Raman interaction with pulse areas of $A_p=\xi_1\pi$ and $A_s=\xi_2\pi$ translates into a $2\pi$ pulse in the MS basis. 
Furthermore, a common phase shift in the pump and Stokes Rabi frequencies in the original basis maps to the same phase shift in $\Omega$ in the MS basis \cite{CpMultiState, RamanGates}. 
Hence, we can use the phases for a composite $2\pi$ pulse (constant-rotation) in order to produce a robust Raman coupling. For example, two such CPs are \cite{TorosovPhaseGate}
\bse
\begin{align}
	&B_0 B_{2\pi/3} C_{0} B_{2\pi/3} B_0, \label{2PiCP}\\
	&B_0 B_{2\pi/5} B_{6\pi/5} B_{2\pi/5} C_{0} B_{2\pi/5} B_{6\pi/5} B_{2\pi/5} B_0,
\end{align}
\ese
where $C=2\pi(1+\epsilon)$ is a nominal $2\pi$ pulse. Therefore, we replace the $P$ and $S$ fields from the Raman coupling in the sequence \eqref{RamanSingleSequence}
with CPs having phases, taken from Eq.~\eqref{2PiCP}. Explicitly, we have
\begin{align}\label{RamanTwoPi}
&\xi_{1,2}\pi \to
\left(\xi_{1,2}\pi/2\right)_0\left(\xi_{1,2}\pi/2\right)_{2\pi/3}\left(\xi_{1,2}\pi\right)_{0} \notag\\
&\qquad\left(\xi_{1,2}\pi/2\right)_{2\pi/3}\left(\xi_{1,2}\pi/2\right)_{0}.
\end{align}
Finally, the $\pi/2$ $Q$ pulse is replaced with one of the constant $\pi/2$ rotations of Eqs.~\eqref{NVV1}, \eqref{NVV2}, or \eqref{BB1}, as we did in the single-Raman-single case.

The excitation profiles for the CP implementation of the Raman-single scenario is compared with the single-pulse case in Fig.~\ref{fig:vsAreaRaman2Wimperis}. 
As in the single-Raman-single approach in Fig.~\ref{fig:vsAreaRaman3Wimperis}, the single-pulse implementation is again sensitive to pulse area errors and works well only for zero errors, $\epsilon=0$ (top left).
The robustness of the method is greatly improved by using composite pulses.
Even the shortest three-pulse CP of Eq.~\eqref{NVV1} delivers a significant enhancement in contrast and error range, and the five-pulse CPs of Eqs~\eqref{NVV2} and \eqref{BB1} offer further improvement.

\section{Implementation with variable-rotation composite sequences}\label{Sec:CompositePulses-variable}

\begin{figure}[tb]
	\includegraphics[width=7cm]{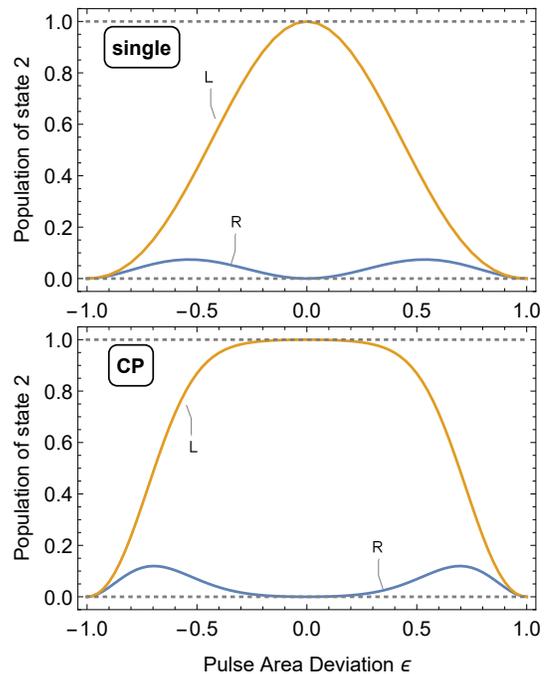}
	\caption{
Population $P_2$ vs pulse area deviation for the single-Raman-single sequence (top frame) and when variable-rotation CPs \eqref{halfpi2} are used to replace the single pulses (bottom frame).
	}
	\label{fig:vsAreaRaman3}
\end{figure}

\begin{figure}[tb]
	\includegraphics[width=7cm]{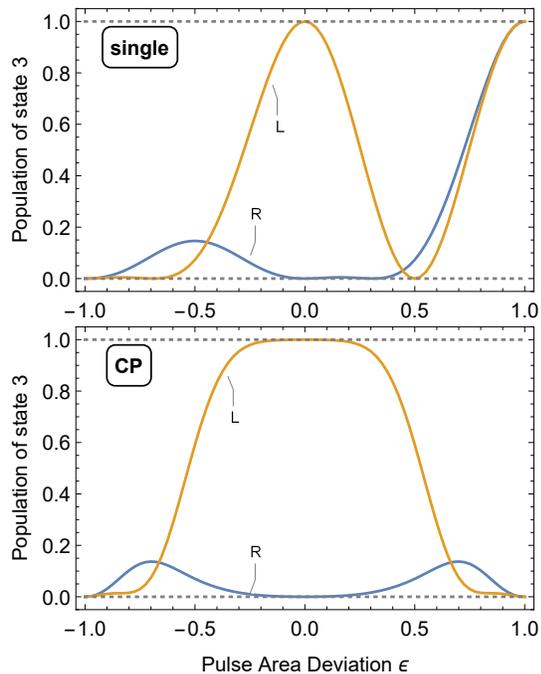}
	\caption{
		Population $P_3$ vs pulse area deviation for the Raman-single sequence (top frame) and when the variable-rotation CP \eqref{halfpi2} is used to replace the single $Q$ pulse and the CP \eqref{2PiCP} is used to replace the $P$ and $S$ pulses in the Raman $[P, iS]$ interaction (bottom frame).
	}
	\label{fig:vsAreaRaman2}
\end{figure}

Up to now, we have shown how to replace the single and Raman pulses with constant composite rotations in order to achieve chiral resolution with high fidelity and robustness. 
However, if we are not aiming for ultra-high fidelity, it turns out that we can also use variable-rotation CPs.
Composite variable rotations are shorter, and hence faster, than constant rotations, which is important if decoherence is present on the time scales of the process.

For instance, in the sequence \eqref{RamanEqxample} we can replace the constituent pulses with the variable-rotation $\pi/2$ CP \cite{TorosovTheta}
\be
(A)_{-\pi/2} (B)_{\pi/4} (A)_{\pi/2} ,\label{halfpi2}
\ee
instead of using the constant-rotation CPs. 
By exploiting the symmetry of the underlying propagators, it can be shown that this CP can also lead to robust chiral resolution, despite the fact that it is of the variable-rotation type. 
To demonstrate this, we can approximate the propagator for the $\pi/2$ $Q$ pulse as
\be
\U_{Q} \approx
 \left[ \begin{array}{ccc}   \frac{e^{i\alpha}}{\sqrt{2}} & 0 & \frac{\pm e^{i\beta}}{\sqrt{2}}  \\ 0 & 1 & 0 \\  \frac{\mp e^{-i\beta}}{\sqrt{2}} & 0 & \frac{e^{-i\alpha}}{\sqrt{2}}  \end{array}\right],
\ee
and for the Raman $[P, S]$ pulse as
\be
\U_{[P,i S]}\approx
\left[ \begin{array}{ccc} \frac{e^{i\gamma_1}}{2} & \frac{e^{i\gamma_4}}{\sqrt{2}} & \frac{e^{-i\gamma_3}}{2} \\ \frac{e^{i\gamma_2}}{\sqrt{2}} & 0 & \frac{-i e^{-i\gamma_2}}{\sqrt{2}}  \\
	\frac{e^{i\gamma_3}}{2} & \frac{i e^{-i\gamma_4}}{\sqrt{2}} & \frac{e^{-i\gamma_1}}{2}  \end{array}\right],
\ee
where the phases $\alpha$, $\beta$ and $\gamma_i$ ($i=1,\ldots, 4$) are parameters which reflect the symmetry properties of the corresponding propagators. By taking the product $\U = \U_{Q} \U_{[P,iS]} \U_{Q}$, after some simple algebra we obtain
\be
U_{21} = \frac{1}{2}e^{-i (\gamma_2+\beta)}\left[\pm i + e^{2\gamma_2+\alpha+\beta} \right].
\ee
Hence, we find that if the condition
\be
2\gamma_2+\alpha+\beta=\pi/2\quad (\text{mod}\ 2\pi)
\ee
is fulfilled, chiral resolution is obtained. It can be shown that by using the CP sequence \eqref{halfpi2}, this condition can be approximately fulfilled over a wide region of pulse area deviation $\epsilon$. 
For this purpose, we need to pay attention to the order of the pulses.
Namely, we need to apply the CP in the same order as in \eqref{halfpi2} when substituting the $Q$ pulses, and in reverse order when substituting the $P$ and $S$ pulses in the $[P, iS]$ Raman interaction.

In Fig.~\ref{fig:vsAreaRaman3} we illustrate the population transfer when this variable-rotation CP is used. 
As seen from the figure, the fidelity is not as high as when constant-rotation CPs are used (Figs.~\ref{fig:vsAreaRaman3Wimperis} and \ref{fig:vsAreaRaman2Wimperis}), but we are using smaller pulse areas and still have much better results than with single pulses.

Similar arguments can be used also for the Raman-single case, where we can substitute the $Q$ pulse with the CP \eqref{halfpi2} (in reversed order) instead of using constant-rotation CPs. 
The performance of this scenario is shown in Fig.~\ref{fig:vsAreaRaman2}. 
We can see from the figure that the fidelity of this method is much greater than for the single-Raman-single case and is even comparable to the constant-rotation approach.

\section{Conclusions}\label{Sec:conclusions}

In this paper we developed two methods for robust high-contrast chirality-dependent population transfer in chiral molecules. 
The methods use sequences of single pulses and Raman pulse pairs applied to a closed-loop three-state system, which has identical properties for the two enantiomers except the opposite signs of one of the couplings.
The closed-loop pattern creates a phase-sensitive interferometric linkage which allows to map the different coupling signs 
onto different populations.

In one of the methods we use a sequence of three interaction steps: a single pulse, a Raman pulse, and another single pulse.
The other method uses only two interaction steps: a Raman pulse, followed by a single pulse.
Both techniques are generalized by replacing the single and Raman pulses with composite pulse sequences of two types: constant and variable rotations.
The composite-pulse implementations achieve very high signal contrast and much greater robustness to experimental errors than single pulses.
Constant rotations feature phase stability and deliver more accurate chiral resolution.
Variable rotations use simpler and shorter (i.e. faster) pulse sequences, which can be important if decoherence is present on the time scale of the process.

\acknowledgments
This work is supported by the European Commission's Horizon-2020 Flagship on Quantum Technologies project 820314 (MicroQC). 
MD acknowledges support from the Independent Research Fond Denmark, the European Commission's
Horizon-2020 FET OPEN Project 766900 (TEQ) and the Villum Foundation.



\begin{thebibliography}{99}
	\bibitem{Noether} E. Noether, Gott. Nachr., 235 (1918).
	
	\bibitem{electroweak}
	A. Bakasov, T.-K. Ha, and M. Quack, J. Chem. Phys. \textbf{109}, 7263 (1998); Err: J. Chem. Phys. \textbf{110}, 6081 (1999);
	M. Quack, J. Stohner, M. Willeke, Annu. Rev. Phys. Chem. \textbf{59}, 741 (2008);
	B. Darqui, C. Stoeffler, A. Shelkovnikov, C. Daussy, A. Amy-Klein, C. Chardonnet, S. Zrig, L. Guy, J. Crassous, P. Soulard, P.	Asselin, T. R. Huet, P. Schwerdtfeger, R. Bast, T. Saue, Chirality \textbf{22}, 870 (2010).
	
	\bibitem{Ahuja} S. Ahuja (editor), \emph{Chiral Separation Methods for Pharmaceutical and Biotechnological Products} (John Wiley \& Sons, 2011).
	
	\bibitem{Chiroptical} N. Berova, P. L. Polavarapu, K. Nakanishi, and R. W. Woody (editors),
	\emph{Comprehensive Chiroptical Spectroscopy: Instrumentation, Methodologies, and Theoretical Simulations}
	(New York, Wiley, 2012)
	
	\bibitem{dichroism} N. Berova and K. Nakanishi, \emph{Circular Dichroism: Principles and Applications} (New York, Wiley, 2000).
	
	\bibitem{Nafie} L. A. Nafie, \emph{Vibrational Optical Activity: Principles and Applications} (Chichester, Wiley, 2011).
	
	\bibitem{NafieStephens} L. A. Nafie, T. A. Keiderling, and P. J. Stephens, J. Am. Chem. Soc. \textbf{98}, 2715 (1976).
	
	\bibitem{Barron} L. D. Barron, \emph{Molecular Light Scattering and Optical Activity} (Cambridge, Cambridge Univ. Press, 2004).
	
	\bibitem{LinearLight} Y. Fujimura, L. Gonz\'{a}lez, K. Hoki, J. Manz, and Y. Ohtsuki,	Chem. Phys. Lett. 306, 1 (1999);
	Y. Fujimura, L. Gonz\'{a}lez, K. Hoki, D. Kr\"{o}ner, J. Manz, and Y. Ohtsuki, Angew. Chem., Int. Ed. \textbf{39}, 4586 (2000);
	K. Hoki, D. Kr\"{o}ner, and J. Manz, Chem. Phys. \textbf{267}, 59 (2001);
	K. Hoki, L. Gonz\'{a}lez, and Y. Fujimura, J. Chem. Phys. \textbf{116}, 8799 (2002);
	L. Gonz\'{a}lez, D. Kr\"{o}ner, and I. R. Sol\'{a}, J. Chem. Phys. \textbf{115}, 2519 (2001);
	D. Kr\"{o}ner, M. F. Shibl, and L. Gonz\'{a}lez, Chem. Phys. Lett. \textbf{372},	242 (2003).
	
	\bibitem{Shapiro} M. Shapiro, E. Frishman, and P. Brumer, Phys. Rev. Lett. \textbf{84}, 1669 (2000); erratum ibid. \textbf{91}, 129902 (2003).
	P. Brumer, E. Frishman and M. Shapiro, Phys. Rev. A \textbf{65}, 015401 (2001).
	D. Gerbasi, M. Shapiro, and P. Brumer, J. Chem. Phys. \textbf{115}, 5349 (2001).
	E. Frishman, M. Shapiro, D. Gerbasi, and P. Brumer, J. Chem. Phys. \textbf{119}, 7237 (2003).
	P. Kr\'{a}l and M. Shapiro, Phys. Rev. Lett. \textbf{87}, 183002 (2001).
	P. Kr\'{a}l, I. Thanopulos, M. Shapiro, and D. Cohen, Phys. Rev. 	Lett. \textbf{90}, 033001 (2003).
	I. Thanopulos, P. Kr\'{a}l, andM. Shapiro, J. Chem. Phys. \textbf{119}, 5105 (2003).
	D. Gerbasi, P. Brumer, I. Thanopulos, P. Kr\'{a}l, and M. Shapiro, J. Chem. Phys. \textbf{120}, 11557 (2004).
	
	\bibitem{STIRAP} N. V. Vitanov, A. A. Rangelov, B. W. Shore and K. Bergmann, Rev. Mod. Phys. \textbf{89}, 015006 (2017).
	
	\bibitem{Hirota}E. Hirota, Proc. Jpn. Acad., Ser. B \textbf{88}, 120 (2012).
	
	\bibitem{M3WM}
	S. R. Domingos, C. P\'{e}rez, and M. Schnell, Annu. Rev. Phys. Chem. \textbf{69}, 499 (2018);
	D. Patterson and J. M. Doyle, Phys. Rev. Lett. \textbf{111}, 023008 (2013);
	D. Patterson, M. Schnell and J. M. Doyle, Nature \textbf{497}, 475 (2013).
	V. A. Shubert, D. Schmitz, D. Patterson, J. M. Doyle, and M. Schnell, Angew. Chem. Int. Ed. \textbf{53}, 1152 (2014);
	V. A. Shubert, D. Schmitz, C. P\'{e}rez, C. Medcraft, A. Krin, S.	R. Domingos, D. Patterson, M. Schnell, J. Phys. Chem. Lett. \textbf{7}, 341 (2016);
	G. G. Brown, B. C. Dian, K. O. Douglass, S. M. Geyer, S. T.	Shipman, and B. H. Pate, Rev. Sci. Instrum. \textbf{79}, 053103 (2008).
	B. G. Park and R. Field, J. Chem. Phys. \textbf{144}, 200901 (2016);	
	C. P\'{e}rez, A. L. Steber, S. R. Domingos, A. Krin, D. Schmitz, and M. Schnell, Angew. Chem. Int. Ed. \textbf{56}, 12512 (2017);	
	S. Eibenberger, J. Doyle, and D. Patterson, Phys. Rev. Lett. \textbf{118}, 123002 (2017).
	
	
	\bibitem{Li2008} Y. Li and C. Bruder, Phys. Rev. A \textbf{77}, 015403 (2008).
	
	\bibitem{Li2019} C. Ye, Q. Zhang, Y.-Y. Chen, and Y. Li, Phys. Rev. A \textbf{100}, 033411 (2019).
	
	\bibitem{LehmannACstark} K. K. Lehmann, arXiv:1501.05282.
	
	\bibitem{Lobsiger2015}S. Lobsiger, C. Perez, L. Evangelisti, K. K. Lehmann, and B. H. Pate, J. Phys. Chem. Lett. \textbf{6}, 196 (2015).			
	
	\bibitem{Lehmann2018} K. K. Lehmann, J. Chem. Phys. \textbf{149}, 094201 (2018).
	
	\bibitem{Koch} M. Leibscher, T. F. Giesen, and C. P. Koch, J. Chem. Phys. \textbf{151}, 014302 (2019).

\bibitem{Goetz2019}
R. E. Goetz, C. P. Koch, and L. Greenman, Phys. Rev. Lett. \textbf{122}, 013204 (2019).
	
	\bibitem{STA}R. G. Unanyan, L. P. Yatsenko, K. Bergmann, and B. W. Shore, Opt. Commun. \textbf{139}, 48 (1997);
	M. Demirplak and S. A. Rice, J. Phys. Chem. A \textbf{107}, 9937	(2003);
	M. V. Berry, J. Phys. A: Math. Theor. \textbf{42}, 365303 (2009);
	X. Chen, I. Lizuain, A. Ruschhaupt, D. Gu\'{e}ry-Odelin, and J. G. Muga, Phys. Rev. Lett. \textbf{105}, 123003 (2010).
	
	
	\bibitem{VitanovDrewsen} N. V. Vitanov and M. Drewsen, Phys. Rev. Lett. \textbf{122}, 173202 (2019).
	
	\bibitem{TorosovDrewsenVitanov} B. T. Torosov,  M. Drewsen, and N. V. Vitanov Phys. Rev. A \textbf{101}, 063401 (2020).
	
	\bibitem{Wu2020} J.-L. Wu, Y. Wang, J.-X. Han, C. Wang, S.-L. Su, Y. Xia, Y. Jiang, J. Song, arXiv:2003.10334.
	
	\bibitem{Wimperis} S. Wimperis, J. Magn. Reson. \textbf{109}, 221 (1994).
		
	
		\bibitem{CpMultiState} G. T. Genov, B. T. Torosov, and N. V. Vitanov, Phys. Rev A \textbf{84}, 063413 (2011)
		
	\bibitem{RamanGates} B. T. Torosov and N. V. Vitanov, arXiv: 2004.12810 (2020).
	
	\bibitem{MajoranaDecomposition} E. Majorana, Nuovo Cimento \textbf{9}, 43 (1932)


	
	\bibitem{MS} J. R. Morris and B. W. Shore, Phys. Rev. A \textbf{27}, 906 (1983).
	
	\bibitem{TorosovPhaseGate}B. T. Torosov and N. V. Vitanov, Phys. Rev. A \textbf{90}, 012341 (2014).
	
	\bibitem{TorosovTheta} B. T. Torosov and N. V. Vitanov, Phys. Rev A \textbf{99}, 013402 (2019).

\end{thebibliography}
\end{document}